\documentclass{emulateapj}

\shorttitle{X-ray Reflection in AGN associated with star formation}
\shortauthors{Wei Yan et al.}

\begin{document}

\title{{\em Chandra} Observations of Excess Fe K$\alpha$ Line Emission in Galaxies with High Star Formation Rates: X-ray Reflection on Galaxy Scales?}

\author{Wei Yan\altaffilmark{1}}
\email{wei.yan.gr@dartmouth.edu}

\author{Ryan C. Hickox\altaffilmark{1}}

\author{Chien-Ting J. Chen\altaffilmark{2}}

\author{Claudio Ricci\altaffilmark{3,4,5}}

\author{Alberto Masini\altaffilmark{6}}

\author{Franz E. Bauer\altaffilmark{7,8,9}}

\author{David M. Alexander\altaffilmark{10}}

\altaffiltext{1}{Department of Physics and Astronomy, Dartmouth College, 6127 Wilder Laboratory, Hanover, NH 03755, USA}

\altaffiltext{2}{Marshall Space Flight Center, Huntsville, AL 35811, USA}

\altaffiltext{3}{N\'ucleo de Astronom\'ia de la Facultad de Ingenier\'ia, Universidad Diego Portales, Av. Ej\'ercito Libertador 441, Santiago, Chile}

\altaffiltext{4}{Kavli Institute for Astronomy and Astrophysics, Peking University, Beijing 100871, China}

\altaffiltext{5}{George Mason University, Department of Physics and Astronomy, MS 3F3, 4400 University Drive, Fairfax, VA 22030, USA}

\altaffiltext{6}{SISSA, Via Bonomea 265, 34151 Trieste, Italy}

\altaffiltext{7}{Instituto de Astrof\'sica, Facultad de F\'isica, Pontificia Universidad Cat\'olica de Chile, Casilla 306, Santiago 22, Chile}

\altaffiltext{8}{Millennium Institute of Astrophysics (MAS), Nuncio Monse\~nor S\'otero Sanz 100, Providencia, Santiago, Chile}

\altaffiltext{9}{Space Science Institute, 4750 Walnut Street, Suite 205, Boulder, CO 80301, USA}

\altaffiltext{10}{Centre for Extragalactic Astronomy, Department of Physics, Durham University, South Road, Durham DH1 3LE, UK}

\received{}
\revised{}
\accepted{}
\published{}

\begin{abstract}

In active galactic nuclei (AGN), fluorescent Fe K$\alpha$ (iron) line emission is generally interpreted as originating from obscuring material around a supermassive black hole (SMBH) on the scale of a few parsecs (pc). However, recent {\it Chandra} studies indicate the existence of iron line emission extending to kpc scales in the host galaxy. The connection between iron line emission and large-scale material can be spatially resolved directly only in nearby galaxies, but could be inferred in more distant AGNs by a connection between line emission and star-forming gas and dust that is more extended than the pc-scale torus. Here we present the results from a stacking analysis and X-ray spectral fitting performed on sources in the {\it Chandra} Deep Field South (CDFS) 7 Ms observations. From the deep stacked spectra, we select sources with stellar mass $\log(M_*/M_\odot)>10$ at $0.5<z<2$, obtaining 25 sources with high infrared luminosity ($\rm SFR_{\rm FIR} \geq 17\;M_{\sun}\; {\rm yr}^{-1}$) and 32 sources below this threshold. We find that the equivalent width of the iron line EW(Fe) is a factor of three higher with 3$\sigma$ significance for high infrared luminosity measured from {\it Herschel} observations, indicating a connection between iron line emission and star-forming material on galaxy scales. We show that there is no significant dependence in EW(Fe) on $M_*$ or X-ray luminosity, suggesting the reflection of AGN X-ray emission over large scales in their host galaxies may be widespread.

\end{abstract}


\section{Introduction}

Through accretion and feedback, Active Galactic Nuclei (AGNs) co-evolve with their host galaxies \citep{alex12bh}. However, the connection between AGN and their host galaxies, in particular the origin of the observed obscuration by gas and dust, remains unclear (e.g., \citealt{hick18}). While type 1 AGNs are luminous in the optical/UV and relatively easy to observe, a majority of AGNs are obscured (e.g., \citealt{hick07abs, mate17obsc, hick18, anan19model}). Obscured AGNs are responsible for the origin of the bulk of the cosmic X-ray background (CXB; \citealt{gill07, trei09, ueda14cxb, aird15nustar, anan19model}). Previous studies also indicate a large population of Compton-thick (CT) AGNs with intrinsic column densities of $N_{\rm H}  \gtrsim 10^{24}\; \rm cm^{-2}$ (e.g., \citealt{lans15nustarqso, ricc15ct, lanz18ct, marc18ct, yan19nustar, geor19ct, carr20}); However, these heavily obscured AGN can be challenging to identify individually, since they are faint in optical and soft X-rays.

One common tracer of heavy AGN obscuration in the X-ray band is the fluorescent Fe K$\alpha$ (iron) line emission at 6.4 keV, a characteristic feature of AGN X-ray spectra. In addition to X-ray spectra of individual sources with the iron line (e.g., \citealt{matt91}), X-ray stacking analyses suggest a prevalence of strong iron line features among CT AGNs (e.g., \citealt{iwas12fe,iwas20cdfs}). More specifically, the obscuration of CT AGNs generates a large equivalent width (EW) of the iron line as high as 1 keV (e.g. \citealt{leve02fe, lama11fe, gand15fe}). For AGNs with high obscuration (e.g., CT AGNs), reflection dominates AGN spectra, and therefore shows characteristics such as a flat spectral shape along with strong iron line emission.

It is generally assumed that iron line photons are produced by reflection from obscuring material in a small-scale ($\sim$ 1--100 pc) region by interacting with the dense circumnuclear `torus’ (e.g., \citealt{shu10fe, ricc14fe}). However, recent {\it Chandra X-ray Observatory} studies of nearby CT AGNs (e.g. \citealt{gand15fe, baue15nustar, mari17fe, fabb17fe, jone20fe}) also discovered kpc-scale diffuse emission of the iron line, suggesting additional reflection of the X-ray emission extending to galactic scales well beyond the parsec-scale `torus'. These direct observations of the extended Fe K$\alpha$ emission are limited to low redshift CT AGNs, for which the kpc scale structures can be resolved by {\it Chandra}. 

Although AGNs at higher redshift cannot currently be resolved in a kpc-scale or smaller in X-rays, multiwavelength observations of more distant AGNs suggest that, the measurement of AGN obscuration can also be notably impacted by gas and dust in host galaxies, instead of only circumnuclear `torus’ region. For example, recent higher resolution studies with the Atacama Large Millimeter Array (ALMA) allow us to probe the central regions of nearby AGNs and image dust as well as molecular gas on small scales, which may be linked to star formation (e.g., \citealt{gall16alma, pere20lirg}). The ALMA observations are also available for some CDFS regions \citep{barg20alma}, suggesting that the interstellar medium (ISM) can produce $N_{\rm H}$ up to CT level in the host galaxy which contributes to AGN obscuration (\citealt{gill14alma, amat20alma}). Therefore, this obscuration by galaxy scale gas may be expected to produce X-ray reflection that could be observed in the fluorescent iron line (e.g., \citealt{circ19host}). In addition to X-ray observations, simulations also suggest that the gas in the galaxy and the circumgalactic medium can contribute significantly to the obscuration along the line of sight of an AGN \citep{treb19}.

In this paper, we explore the connection between integrated Fe K$\alpha$ line emission and galaxy properties in distant AGNs (0.5 $<$ z $<$ 2.0) with deep extragalactic survey data. After conducting a stacking analysis and detecting the iron line in average X-ray spectra, we measure the dependence of the equivalent width of the iron line [EW(Fe)], on estimated obscuring column density ($N_{\rm H}$) as well as its connection to galactic properties derived from multi-wavelength surveys. The paper is organized as follows: Section 2 details the X-ray data analysis, and multi-wavelength surveys are discussed in Section 3. We discuss our results in Section 4 and summarize the paper in Section 5. Throughout the paper, we assume a $\Lambda$CDM cosmology with $H_0=69.6 \rm {km\;s}^{-1}{\rm Mpc}^{-1}$, ${\rm \Omega}_{\rm M}=0.286$ and ${\rm \Omega}_{\rm \Lambda}=0.714$ \citep{wrig06coscal}.

\begin{figure}
\includegraphics[width=90mm]{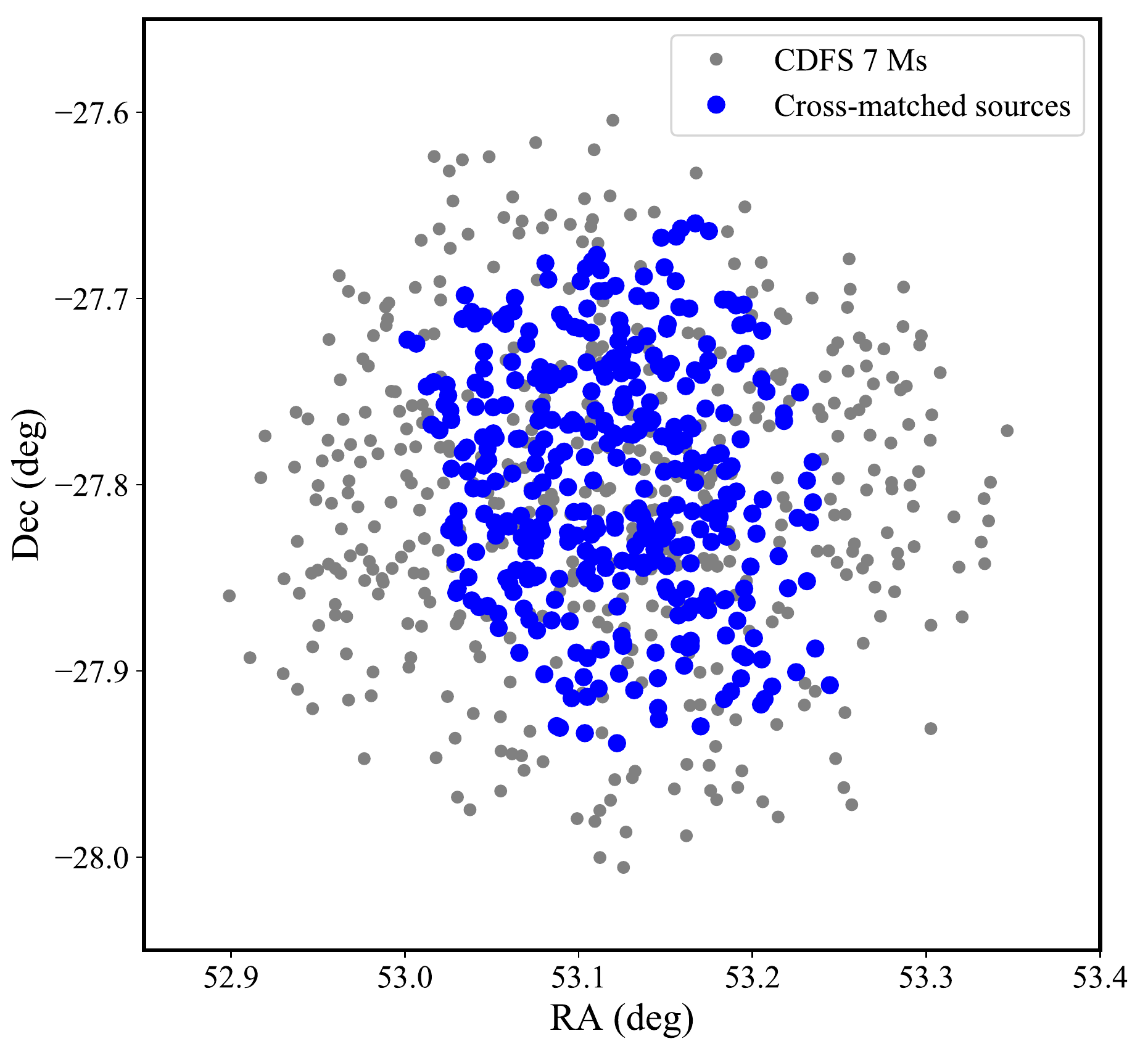}
\caption{The sky map of all sources in the CDFS 7 Ms catalog (grey) and the cross-match sources (see Section 3 for detials) with {\it Herschel} observations and optical SED fitting in \citet{sant15sed} (blue).  \label{fig:sky3}}
\end{figure}

\section{X-ray Data Analysis}

We obtain the CDFS 7 Ms X-ray observations from {\it Chandra} archive and extract spectra of all sources to derive the average EW(Fe) in X-ray emission. This deep survey contains 102 observations in a total area of 484.2 $ \rm arcmin^2$, including 1055 classified AGNs and galaxies with available spectroscopic redshifts taken from $\sim$ 30 public catalogs and photometric redshifts collected from 6 catalogs (see \citealt{luo17chandra} for details). 986 sources in CDFS 7 Ms catalog have available redshifts, including 653 secure spectroscopic redshifts and 333 photometric redshifts of high quality with outliers as low as 1.1\% (see Figure 10 in \citealt{luo17chandra}). For full-band detections in the energy range 0.5--7.0 keV, X-ray band ratios (Br, defined as the ratio of count rates between the hard-band 2.0--7.0 keV and soft-band 0.5--7.0 keV; \citealt{luo17chandra}) are calculated using the Bayesian code $BEHR$ \citep{park06behr}. The median number of counts in the spectra in the 0.5--7.0 keV band is 98.9. For the rest of the sources, band ratios are adopted from the mode values of the band-ratio probability density distributions for best-guess estimates instead of upper or lower limits \citep{luo17chandra}. From this catalog, we also utilize the intrinsic X-ray luminosities of all sources in the rest frame 0.5--7.0 keV of the observed full-band, soft-band, or hard-band. By adopting a Galactic column density of $N_{\rm H} = 8.8\times10^{19} \; {\rm cm}^{-2}$ along the line of sight to the CDFS (e.g., \citealt{star92}) and assuming fixed photon index values, \citet{luo17chandra} estimate intrinsic absorption $N_{\rm H}$. We adopt these values as well as the corrected rest-frame 0.5--7.0 keV  X-ray luminosity ($L_{X}$) from the catalog with a cleaned net exposure time around 6.727 Ms for on-axis sources. Although most sources have relatively few numbers of counts, these low-count spectra can be stacked to obtain high signal-to-noise (SNR) average spectra. 

\begin{figure*}[t]
\epsscale{1.1}
\plotone{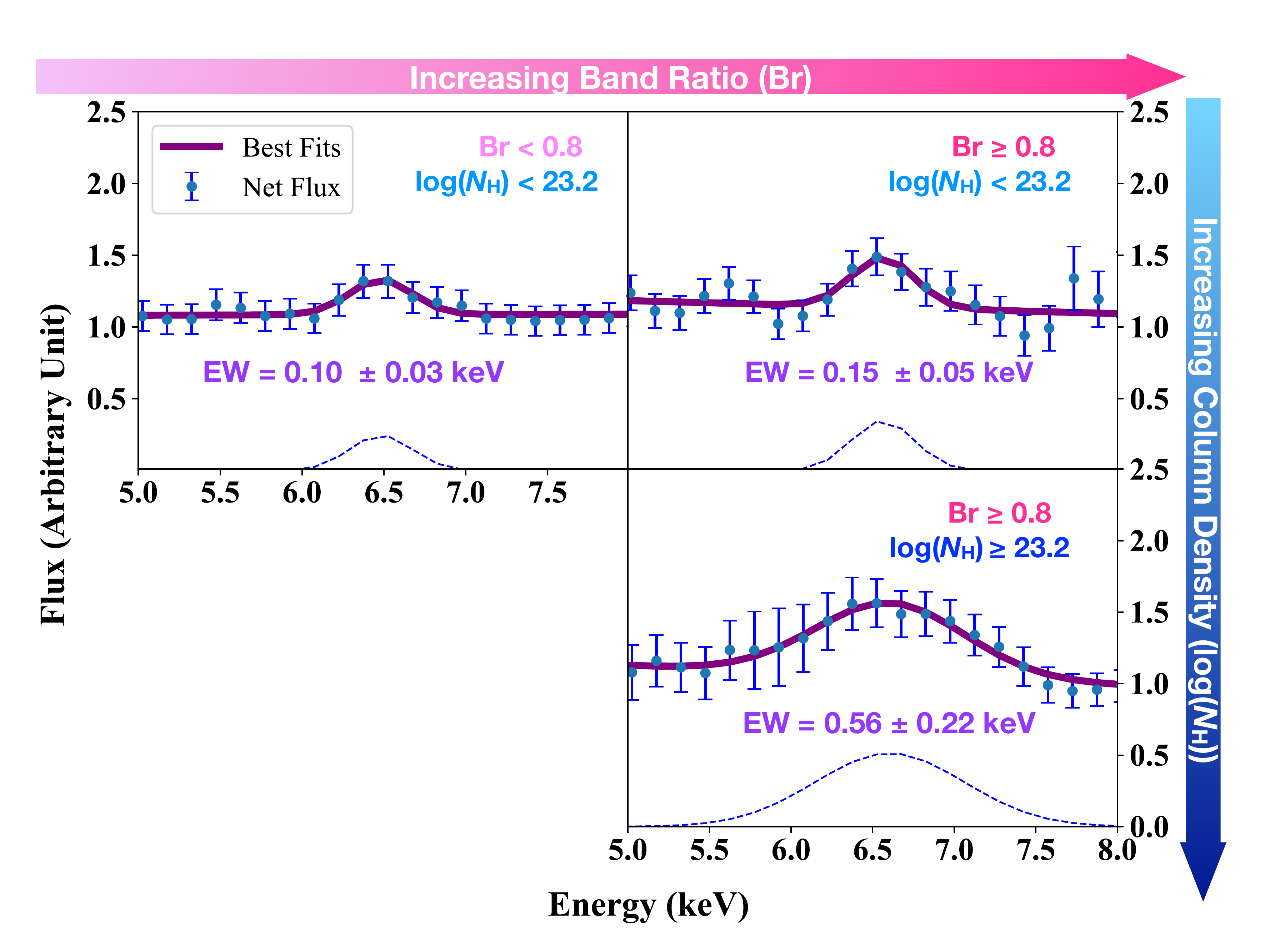}
\caption{Stacked spectra in grouped CDFS 7 Ms sources, divided by $N_{\rm H}$ and band ratio (Br) from the catalog (98 sources with low Br and $N_{\rm H}$, 69 sources with high Br and low $N_{\rm H}$, and 88 sources with high Br and $N_{\rm H}$). The average spectra show clear Fe K$\alpha$ lines at 6.4 keV with EW as high as 0.56 keV. The gradient blue represents the obscuration level (higher $N_{\rm H}$ with darker blue), while the gradient pink represents the value of Br indicating the obscuration level (higher Br and more obscured with darker pink). Blue dots and purple lines show the stacked fluxes in each energy bin and the best fits, respectively. Dashed lines show the Gaussian component of the best fits. The critical values of Br = 0.8 and $\log N_{\rm H} = 23.2$ are approximately the median values of selected sources. The group with higher Br and higher $N_{\rm H}$ shows the highest EW(Fe).  \label{fig:br}}
\end{figure*}

\subsection{Stacking Analysis}

To perform the stacking analysis, we first use $CIAO$ to extract the source and background X-ray spectra, Auxiliary Response Files (ARFs), and combined Redistribution Matrix Files (RMFs) of individual sources from each observation. We set the source region as a circle around each source. Since {\it Chandra}'s point-spread function (PSF) changes with off-axis angles, larger extraction regions are adopted for sources at large off-axis angles in order to optimize SNR (e.g., \citealt{capp17cosmos}). For each source, we use $r_{90}$\footnote{$\it{Chandra}$ Proposer’s Observatory Guide ( POG), ver. 23.0, Fig. 4.13, available at https://cxc.harvard.edu/proposer/POG.}as a source region radius which represents the 90\% encircled energy radius [$r_{90} \sim 1\arcsec + 10\arcsec(\theta/10\arcmin)^2$, where $\theta$ is the off-axis angle]. We then set the background region as an annulus around the source with radii between $2r_{90}$ and $3r_{90}$, excluding all other sources within this region.

Since one single spectrum of one source usually has very low SNR, in most cases the iron line emission is not strong enough to be separated from the noise. Therefore, we first combine the source and background spectra from individual observations of each source. Then, for every source, we subtract the scaled background from the source region and use \textsc{sherpa} to model the combined spectrum \citep{free01sherpa}. To begin with, we fit the continuum of the spectrum with an absorbed power law model modeled by $\it{xszphabs}$. In the continuum fitting we exclude the energy range between rest-frame 5.5 and 7.5 keV to avoid the effects of the iron emission line. Using the power-law fit to the continuum, we then group each spectrum with a minimum of 20 photons in order to obtain the ratio of the observed flux to the best-fit continuum model in each energy bin. Thus for each source we obtain a measurement of the strength of the iron line emission relative to the continuum. We use the ratio relative to the continuum rather than directly averaging the spectra, because we are mainly interested in the properties of the iron emission line. Since the shape of the continuum varies for different sources, if we were to average the spectra directly, this would introduce significant scatter in the measurement of EW(Fe). 

In order to convert the obtained ratios from observed frame energies to a uniform rest frame, we first correct the energies for redshift. Then we interpolate the ratios to rest-frame energies between 1 and 15 keV with a uniform bin size of 0.15 keV. To further determine the uncertainties in the rest frame, we propagate the Poisson counting errors to each bin in rest-frame energy, taking the change of bin size from the grouped bin size in the observed frame to the uniform rest-frame energy bin size (0.15 keV) into consideration.

We aim to use AGN characteristics  (e.g., obscuration and band ratio) to evaluate the validation of EW(Fe) obtained from our stacking analysis described above. Therefore, we use $N_{\rm H}$ and Br to divide the catalog of CDFS sources into groups. We obtain the average spectra of sources in each group in order to look for the trend between EW(Fe) and these two parameters.

There are 962 sources with available $N_{\rm H}$ estimates as well as Br values from the catalog. In order to focus on the relation between iron line emission and obscuration, we require sources with significant SNR around the Fe K$\alpha$ line. We therefore only include sources detection in the hard band (2--7 keV at observed frame), eliminating the sources in the lowest quartile in the number of hard band net counts (net counts $>$ 77). We further require a SNR $>$ 5 in the hard band, yielding a total of 264 selected sources. Some sources with weak detections show extreme flux ratios after X-ray spectral fitting. The quality of their fits is poor due to the lack of photons in continuous energy bins. This prevents us from obtaining a good estimate of the continuum, which could introduce large biases during our stacking procedure. To remove these outliers, we calculate the average flux ratio between 4.8--8 keV at rest frame for every source and derive the distribution. We then exclude those sources beyond 3$\sigma$ from the center value. After the selection, we obtain 255 sources in total.

We acknowledge that removing sources with low numbers of counts could introduce a selection bias in our analysis. However, many of the low-count sources are heavily obscured AGN, whose intrinsic X-ray luminosities maybe underestimated by as much as a factor of 2 \citep{lamb20cdfs}. By restricting our analysis to brighter sources whose intrinsic luminosity is better constrained, we maximize our SNR and avoid uncertainty in comparing intrinsic $L_{\rm X}$ between high and low SFR sources. Within our sample of 264, we are confident that the intrinsic $L_{\rm X}$ distributions of the high and low SFR sources are consistent with each other, so that SFR is the dominant factor in the observed difference in EW(Fe).

We further divide these 255 sources into three groups based on their $N_{\rm H}$ and Br values \citep{luo17chandra}. Each group has a similar enough number of sources so that we are able to obtain the average stacking spectrum with comparable SNR.

Group 1 sources have 98 sources with Br $<$ 0.8 and $N_{\rm H} < 10^{23.2}\; \rm cm^{-2}$; Group 2 have 69 sources with Br $\geq$ 0.8 and $N_{\rm H} < 10^{23.2}\; \rm cm^{-2}$; and Group 3 have 88 sources Br $\geq$ 0.8 and $N_{\rm H} \geq 10^{23.2}\; \rm cm^{-2}$. The critical value 0.8 is approximately the median Br value of the selected sources, and $10^{23.2}\; \rm cm^{-2}$ is also approximately the median $N_{\rm H}$ value for source with Br $\geq$ 0.8. Here we assume EW(Fe) is an independent indicator of nuclear obscuration, and the fitting of the average spectra of each group is shown in Figure \ref{fig:br}.

We fit the continuum and the emission at $6.4\;$keV with a linear component and a Gaussian line component, respectively. From the fitting, we obtain the slope of the power law and derive EW(Fe). The power law is fitted to a flat line with small scatter around 1, since the continuum of the average spectrum shows the ratio of the observed flux to the continuum model. We find the group with higher Br and higher $N_{\rm H}$ shows the highest EW(Fe). To estimate the uncertainty of EW(Fe) in each stacked spectrum, we perform bootstrap resampling, selecting a random subset of the sources (with replacement) and re-computing the average spectrum. We perform the bootstrap 500 times and compute the dispersion of the stacked spectrum in each energy bin.

\begin{figure}
\includegraphics[width=85mm]{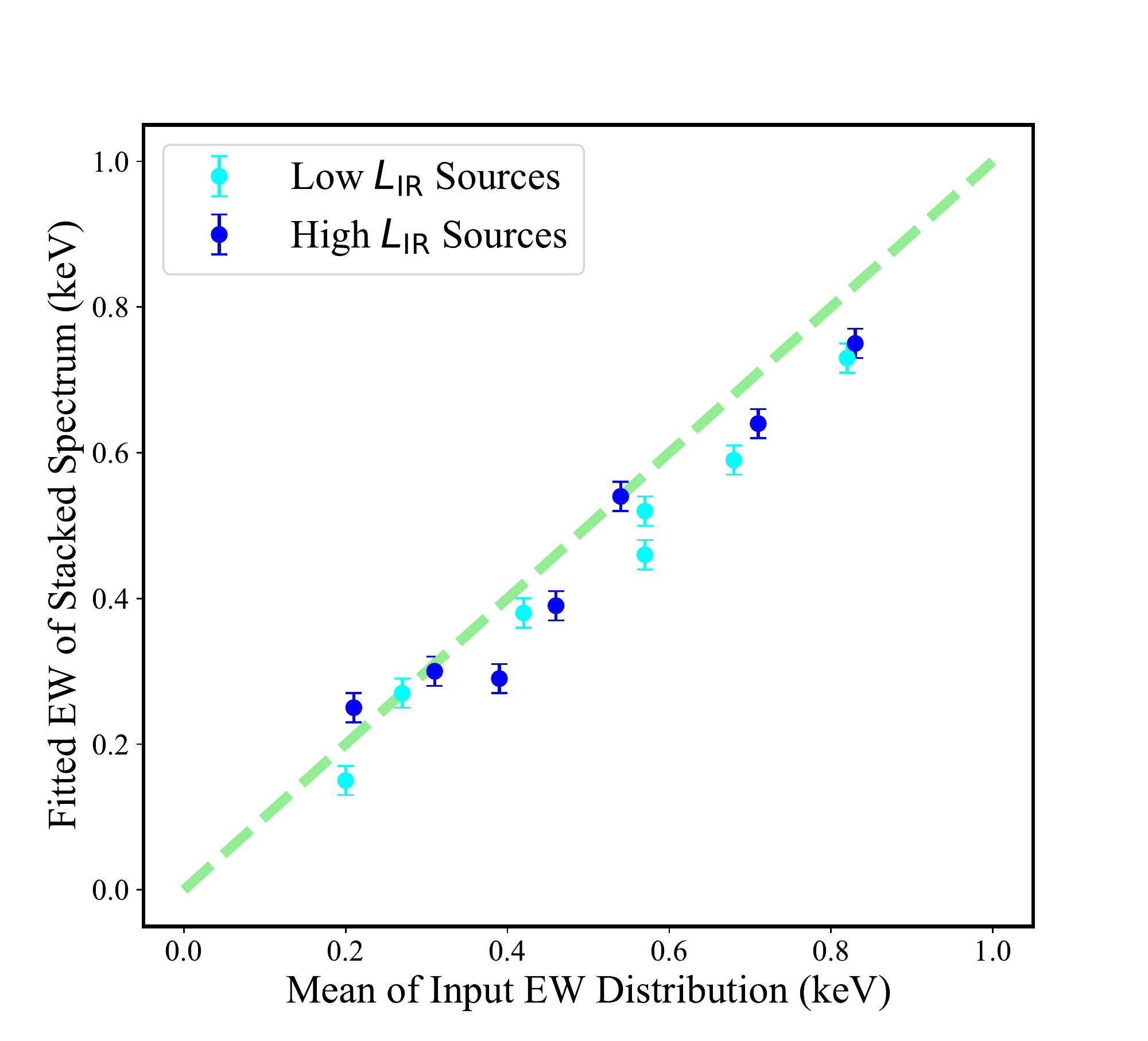}
\caption{Comparison between output and input EW(Fe) values in simulations. The simulated spectra follow the same distributions of secure spectroscopic redshift and photon index of our selected sample. Blue points are sources with high $L_{\rm IR}$ and cyan points are those with low $L_{\rm IR}$ (sample selection see Section 3). EW(Fe) value of individual simulated spectrum is drawn randomly from Gaussian distribution with means from 0.2 to 0.8 keV. The green dashed line shows a 1 to a relation. After conducting the stacking analysis, the fitted EW(Fe) of the stacked spectrum is slightly lower but still consistent with the input mean within $\sim$ 80\%, regardless the FWHM value. \label{fig:sim}}
\end{figure}


\subsection{Verification}

To verify our stacking techniques, we perform the same X-ray stacking procedure with simulated spectra in {\it sherpa} to test our fitted method. By using spectral models with the same spectroscopic redshift and gamma distribution of our selected sample (see sample selection in Section 3), we simulate spectra with a randomly assigned input EW(Fe) from a Gaussian distribution. We conduct the same stacking analysis and convert the ratio of observed flux to continuum model to the uniform rest frame. From there, we fit the average spectra and obtain fitted values of EW(Fe). We then repeat this process using different Gaussian distributions with mean value varying from 0.2 keV to 0.8 keV. In our simulations, we also set different full width at half maximum (FWHM) energy values range from 0.1 keV to 1.0 keV at the rest frame. As a result, the average simulated spectra have consistent FWHM outputs, showing similar broad shape as our stacked spectra. These obtained EW(Fe) outputs are generally consistent with the average input values (Figure~\ref{fig:sim}). We acknowledge that for both low and $L_{\rm IR}$ groups, the outputs tend to be slightly smaller than the inputs as the EW(Fe) value increases. This suggests that, although we might slightly underestimate the strength of iron line emissions, the intrinsic difference of the observed spectra between two $L_{\rm IR}$ groups is not affected by our fitting procedure.

We notice that in order to produce a spectral shape similar to the observations, the dispersion of the simulated spectra has to be set at a relatively high level ($\sigma \sim 0.3$ keV at rest frame). This suggests the presence of a broad Fe K$\alpha$ emission line in our stacked spectrum. Both observed stacked spectra of low and high $L_{\rm IR}$ sources show comparable broadening. The breadth of this emission feature cannot be explained by the ACIS instrumental energy resolution, intrinsic velocity broadening, or redshift errors, all of which produce $\sigma\lesssim0.1$ keV. Therefore the origin of the precise shape of the average iron line is uncertain and warrants further study. However, our simulations show that the measurement of the strength of the Fe line emission as parameterized by EW(Fe) is unaffected by this unusually large line width; our fitted EW(Fe) value in the stacked spectrum is always consistent with the average of all input spectra, regardless the chosen FWHM values.

\begin{figure*}[t]
\epsscale{1.1}
\plotone{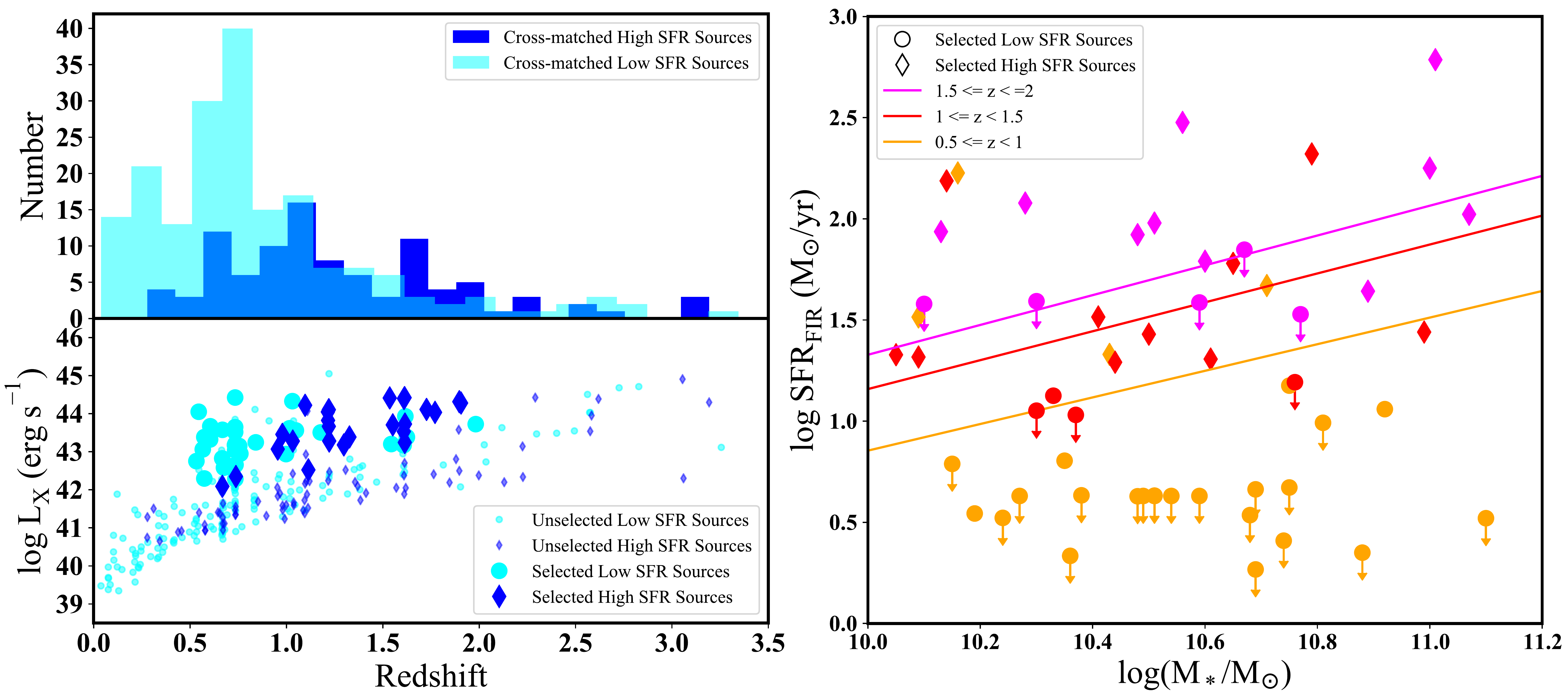}
\caption{$\it{Left}$: Redshifts and X-ray luminosities distribution of all 289 cross-matched sources. These cross-matched sources are separated into two groups based on their $\rm SFR_{\rm FIR}$: high SFR group in blue diamonds with $\rm SFR_{\rm FIR} \geq 17\;M_{\sun}\; {\rm yr}^{-1}$, and low SFR group in cyan dots with $\rm SFR_{\rm FIR} < 17\;M_{\sun}\; {\rm yr}^{-1}$. Far-IR values are adopted from \citet{mull12fir}. We further select those with z = 0.5--2.0 and good SNR (see details in Section 3) from the cross-matched sources to perform stacking analysis. $\it{Right}$: $\rm SFR_{\rm FIR}$ versus $M_*$ of our selected sources. Stellar masses are adopted from \citet{sant15sed}. Low SFR and high SFR groups are shown as big dots and diamonds, respectively. Solid lines are main-sequence relations at different redshift bins (orange: z = 0.5--1, red: z = 1--1.5, magenta: z = 1.5--2) adopted from \citet{spea14ms}. The widths of these relations are taken as $\pm$ 0.2 dex scatters. SFR affects EW(Fe) more significantly than sSFR. \label{fig:sfr-z}}
\end{figure*}

\section{Multiwavelength Surveys and Infrared Luminosity}

The CDFS field is one of the most intensively studied multi-wavelength deep survey regions across the entire sky. Optical and infrared (IR) counterparts are identified and provide basic information such as optical fluxes, spectra, or morphologies (e.g., \citealt{mull12fir, sant15sed}). Insecure spectroscopic redshifts in the CDFS 7 Ms catalog may introduce as large as 15\% uncertainties and photometric redshifts may have even larger uncertainties. These errors in redshifts can significantly affect the accuracy of our EW(Fe) measurements. Therefore, here we only select sources in the CDFS 7 Ms catalog with secure spectroscopic redshifts, and then cross-match these sources with spectral energy distribution (SED) fitting results in the CANDELS/GOODS-South catalog from \citet{sant15sed}, which excludes sources with bad photometry (see \citealt{guo13candels, gala13candels}). From the far-IR observations in the {\it Herschel} catalog from the \citet{mull12fir}, we adopt far-IR luminosity ($L_{\rm FIR}$) and estimate far-IR SFR (star formation rate, hereafter $\rm{SFR}_{\rm FIR}$) values. In \citet{mull12fir}, 100 $\mu$m and 160 $\mu$m flux densities for AGNs at z = 0.5--1.5 and z = 1.5--3, respectively, are used to estimate $L_{\rm FIR}$ using the empirical SED templates of \citet{mull11agnsed}. $\rm{SFR}_{\rm FIR}$ estimates are calculated from the host galaxy $L_{\rm FIR}$ using the prescription outlined in \citealt{kenn98araa}, assuming a \citealt{salp55lf} IMF and solar luminosity ($L_{\sun}$) as $3.8\times10^{33}\; \rm ergs \; s^{-1}$, i.e.,

\begin{equation}
\frac{\rm SFR_{\rm FIR}}{M_{\sun}\; {\rm yr}^{-1}}=1.7\times10^{-10}\;\frac{L_{\rm FIR}}{L_{\sun}}
\end{equation}

We exclude sources on the edge of the {\it Herschel} field with low SNR and weak detection ($L_{\rm FIR} \sim 10^{-10}\;L_{\sun}$), resulting in the cross-matched sources shown as blue dots in Figure \ref{fig:sky3}. We only focus on sources with redshifts between 0.5 to 2.0, which comprises the bulk of the sources in the distribution of far IR luminosity $L_{\rm FIR}$. Additionally, at higher redshifts, the galactic properties (e.g., SFR; stellar mass, or $M_*$) of certain galaxies might be particularly uncertain due to possible issues related to the SED fitting. \citet{sant15sed} found that the stellar masses of young galaxies (age $ < $ 100 Myr), in particular redshift ranges (e.g., 2.2 $< z <$ 2.4), can be significantly overestimated (by up to a factor of 10 for age $ < $ 20 Myr sources) if the nebular contribution is ignored. Following \citet{sant15sed}, we adopt the median values of $M_*$ and SFR (right panel in Figure~\ref{fig:sfr-z}) as computed using different methods by five separate teams, which show consistent values obtained with independent templates (labeled as $2a\tau$, $6a\tau$, $11a\tau$, $13a\tau$ and $14a$, see \citealt{sant15sed, yang17}).

\begin{figure*}[t]
\epsscale{1.1}
\plotone{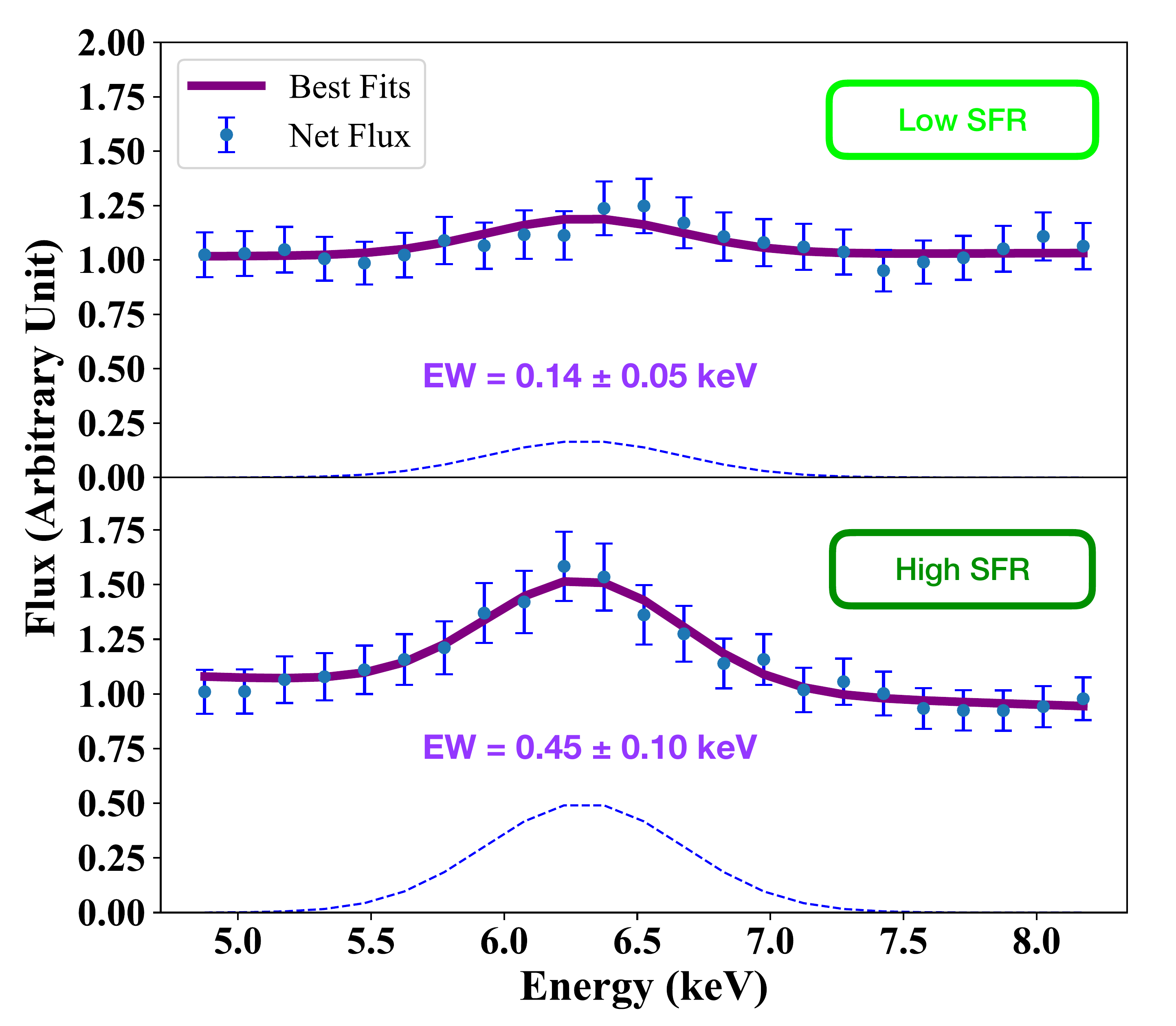}
\caption{Stacked spectra between 5--8 keV of 32 sources with low SFR (top panel) and 25 sources with high SFR (bottom panel) in far-IR. Net flux is shown in blue with errors derived from bootstrap resampling. Solid lines show the best fits and dashed lines show the Gaussian component of the best fits. The higher SFR spectrum shows a much stronger iron line, with the EW(Fe) increasing by a factor close to 3. \label{fig:mainfit}}
\end{figure*}

\section{Results}
We now consider the connection between galactic properties and EW(Fe), following the X-ray stacking procedure laid out in Section 2. For the 289 cross-matched sources with secure spectroscopic redshifts between 0.5 and 2.0, we further select sources with detected net hard band (2.0--7.0 keV at observed frame) counts over 77, which removes the faintest 25\% of sources from the highest 75\%, in order to obtain average spectra with clear signals. We also exclude sources with average flux ratios above the 3$\sigma$ limit between 4.8--8.0 keV at rest frame, following the same strategy in the last section.

We then divide the sources into two groups based on the thresholds of ${\rm SFR}_{\rm FIR}$: 26 sources with $\rm SFR_{\rm FIR} \geq 17\;M_{\sun}\; {\rm yr}^{-1}$ so called ``high SFR" shown as large blue dimonds and blue columns in Figure \ref{fig:sfr-z}, and 40 sources with $\rm SFR_{\rm FIR} < 17\;M_{\sun}\; {\rm yr}^{-1}$, ``low SFR" shown as cyan columns in Figure \ref{fig:sfr-z} for comparison). To confirm that SFR is the only different parameter between the two groups, we compare the distributions of $M_*$  and $L_X$ of both groups by conducting a 2D KS test. Although we obtain a p-value of 44.5\% and cannot reject the similarity between these two distributions, we find that the low SFR group has more low $M_*$ sources than the high SFR group, consistent with expectations from the star-forming ``main sequence" (MS, e.g., \citealt{elba11ms_aph}). To be sure that SFR is the dominant difference in two groups, we only consider sources with $\log(M_*/M_\odot)>10$, which leaves 25 sources with ``high SFR" and 32 sources with ``low SFR". 

As a further check, we confirm that this result is not dominated by the brightest X-ray sources (with the smallest corresponding uncertainties on continuum modeling) or any outliers. For this, we introduce a normalization scheme that provides significant additional weight to faint sources, while avoiding excessive scatter due to larger Poisson errors among the fainter sources. 

To this end, we average the flux to continuum ratios for each spectrum as described above, but weight sources by the quantity $(\log{F_{\rm hb}}-0.99*\log{F_{\rm hb}}_{\;\rm min})^3$, where $F_{hb}$ is the hard-band flux listed in the CDFS 7 Ms catalog for each source, and $F_{\rm hb\;min}$ is the faintest hard band flux for sources included in our stacking. This weighting scheme is designed so that the faintest 50\% of the sources contributed at least 25\% to the final stacked spectrum. With this weighting procedure, the results are unchanged, with a difference in EW(Fe) between high and low $\rm SFR_{\rm FIR}$ sources of approximately a factor of 3. We perform bootstrapping to estimate EW(Fe) uncertainties, and confirm that the EW(Fe) difference has a significance of 3$\sigma$. Our stacked spectra represent the average EW(Fe) of selected sources, which is comparable to previous EW(Fe) measurements of individual sources in COSMOS field (e.g., two sources with EW $\sim$ 0.15 keV and 0.3 keV, respectively, in \citealt{iwas12fe}).

We notice that sources in the low SFR group have lower redshift than those in the high SFR group (left panel in Figure~\ref{fig:sfr-z}). In order to confirm whether redshift dominates the EW(Fe) values in these two groups, we perform the same analysis on sources in smaller redshift bins (e.g., 0.5 $\leq$ z $\leq$ 1.5, 1 $\leq$ z $\leq$ 1.5). As a result, the significant difference between EW(Fe) values in two groups remains as 2--3 $\sigma$, regardless of the redshift range. We also compare the $N_{\rm H}$ distribution of the two groups based on the intrinsic values obtained by X-ray spectral fits in \citep{luo17chandra}. With a p-value of 15.7\% and comparable mean ($\log N_{\rm H}$ = 22.89 and 22.55, respectively), we cannot rule out the similarity between these two distributions. We notice that the high-SFR group contains a few sources with relatively low $N_{\rm H}$. Since we exclude sources with low counts which may include heavily obscured AGNs with high SFR, the X-ray detection here is more likely from unobscured AGNs rather than star forming activity.

Moreover, we divide all the selected sources based on their specific SFR (sSFR, defined as SFR/$M_*$) as well. We separate sources by the MS sSFR in each redshift bin (Figure~\ref{fig:sfr-z}; \citealt{spea14ms}). We adopt an intrinsic scatter in the MS relationships of $\pm$ 0.2 dex. The high sSFR group of sources with sSFR above the MS relation has an EW(Fe) = 0.36 $\pm$ 0.11 keV, and the low sSFR group of the rest shows an EW(Fe) = 0.25 $\pm$ 0.08 keV. The difference between two sSFR groups is smaller than SFR groups, which indicates that SFR affects EW(Fe) much more than sSFR. We note that the upper limits of three FIR-undetected sources are slightly above the MS relation. We have verified that including these specific sources in either group does not significantly affect the stacked average spectra. Therefore to avoid ambiguity do not include these three sources in either sSFR group.

Therefore, as SFR appears to be the only significantly different galactic parameter related to EW(Fe), we argue that the strength of iron line EW(Fe) has a correlation with SFR of the host galaxies. The increase EW(Fe) suggests that instead of relating to heavy absorption in torus-scale region, the iron line emission is related to materials associated with larger galactic-scale star formation.
 

One direct interpretation is that the observed Fe K$\alpha$ line is extended to galactic scale ($\sim$kpc) due to reflection caused by galactic star-forming clouds. Previous studies (e.g., \citealt{baue15nustar, fabb17fe, jone20fe}) have detected extended Fe K$\alpha$ emission ($\sim$kpc scale) in a few nearby galaxies. Our results suggest that extended Fe K$\alpha$ emission may be a common feature for distant galaxies. 

While these results indicate X-ray reflection beyond the parsec-scale torus, they do not place constraints on the extent of reflecting material. The fueling of AGN is known to be strongly correlated with star formation in the central kpc of the galaxy (e.g., \citealt{dima12agnsf, esqu14agnsf_aph}), so it is possible that the reflection is associated with these nuclear starbursts. High resolution observations, for example with the ALMA or the {\em James Webb Space Telescope}, may be able to measure the distribution of star formation in these systems and better constrain the physical extent of the X-ray reflecting material.

\section{Conclusion}

We have performed a stacking analysis to compute the average X-ray spectra of AGN with redshifts between 0.5 and 2.0 in the CDFS 7 Ms catalog. Dividing the sources between those that are more and less luminous in the FIR as determined by {\it Herschel} observations, we find a clear connection between FIR luminosity (and thus galactic SFR) and the strength of the Fe K$\alpha$ line. We interpret this relation as being due to reflection of nuclear emission by the star-forming gas that is typically distributed on galaxy scales. This observed relationship clearly indicates that nuclear emission from AGN can be strongly affected by gas and dust in the host galaxy (e.g., \citealt{hick18, circ19host, amat20alma})

Considering this result in the context of previous discoveries of extended iron line limited in nearby CT AGNs (e.g. \citealt{gand15fe, baue15nustar, mari17fe, fabb17fe}), our analysis suggests that Fe line reflection on galaxy scales may be a widespread phenomenon in AGN at moderate to high redshifts. Furthermore, the presence of extended Fe K$\alpha$ line emission in distant AGN presents a challenge in modeling of their X-ray spectra, which is often carried out assuming absorption and reflection as the result of a small (pc-scale) torus rather than galaxy-scale material (e.g., \citealt{netz15unified, brig17ct, balo18borus, pana19torus}). Future observations with more sensitive (e.g., {\it Athena}) or higher-resolution X-ray observations (e.g., {\it Lynx}), may be able to directly measure the relation between EW(Fe) and galactic SFR for individual AGN at larger redshift, which may shed a light on the nature of connection between black hole growth and galaxies over cosmic time.

\acknowledgments

\begin{acknowledgments}

We thank the anonymous referee and acknowledge support from the National Science Foundation through CAREER award number 1554584 and by Dartmouth Fellowship. We also acknowledge support from the Fondecyt Iniciacion grant 11190831 (C.R.), CONICYT grants CATA-Basal AFB-170002 (F.E.B.), FONDECYT Regular \#1190818 (F.E.B.) and \#1200495 (F.E.B.), FONDECYT iniciacion \#11190831; and
Chile's Ministry of Economy, Development, and Tourism's Millennium Science Initiative through grant IC120009, awarded to The Millennium Institute of Astrophysics, MAS (FEB).

\end{acknowledgments}

 \bibliographystyle{apj}

\begin{thebibliography}{}
\expandafter\ifx\csname natexlab\endcsname\relax\def\natexlab#1{#1}\fi

\bibitem[{{Aird} {et~al.}(2015){Aird}, {Alexander}, {Ballantyne}, {Civano},
  {Del-Moro}, {Hickox}, {Lansbury}, {Mullaney}, {Bauer}, {Brandt}, {Comastri},
  {Fabian}, {Gandhi}, {Harrison}, {Luo}, {Stern}, {Treister}, {Zappacosta},
  {Ajello}, {Assef}, {Balokovi{\'c}}, {Boggs}, {Brightman}, {Christensen},
  {Craig}, {Elvis}, {Forster}, {Grefenstette}, {Hailey}, {Koss}, {LaMassa},
  {Madsen}, {Puccetti}, {Saez}, {Urry}, {Wik}, \& {Zhang}}]{aird15nustar}
{Aird}, J., {Alexander}, D.~M., {Ballantyne}, D.~R., {et~al.} 2015, \apj, 815,
  66

\bibitem[{{Alexander} \& {Hickox}(2012)}]{alex12bh}
{Alexander}, D.~M., \& {Hickox}, R.~C. 2012, \nar, 56, 93

\bibitem[{{Ananna} {et~al.}(2019){Ananna}, {Treister}, {Urry}, {Ricci},
  {Kirkpatrick}, {LaMassa}, {Buchner}, {Civano}, {Tremmel}, \&
  {Marchesi}}]{anan19model}
{Ananna}, T.~T., {Treister}, E., {Urry}, C.~M., {et~al.} 2019, \apj, 871, 240

\bibitem[{{Balokovi{\'c}} {et~al.}(2018){Balokovi{\'c}}, {Brightman},
  {Harrison}, {Comastri}, {Ricci}, {Buchner}, {Gandhi}, {Farrah}, \&
  {Stern}}]{balo18borus}
{Balokovi{\'c}}, M., {Brightman}, M., {Harrison}, F.~A., {et~al.} 2018, \apj,
  854, 42

\bibitem[{{Barger} {et~al.}(2020){Barger}, {Cowie}, {Bauer}, \&
  {Gonzalez-Lopez}}]{barg20alma}
{Barger}, A., {Cowie}, L., {Bauer}, F., \& {Gonzalez-Lopez}, J. 2020, in
  American Astronomical Society Meeting Abstracts, Vol. 236, American
  Astronomical Society Meeting Abstracts \#236, 122.01

\bibitem[{{Bauer} {et~al.}(2015){Bauer}, {Ar{\'e}valo}, {Walton}, {Koss},
  {Puccetti}, {Gandhi}, {Stern}, {Alexander}, {Balokovi{\'c}}, {Boggs},
  {Brandt}, {Brightman}, {Christensen}, {Comastri}, {Craig}, {Del Moro},
  {Hailey}, {Harrison}, {Hickox}, {Luo}, {Markwardt}, {Marinucci}, {Matt},
  {Rigby}, {Rivers}, {Saez}, {Treister}, {Urry}, \& {Zhang}}]{baue15nustar}
{Bauer}, F.~E., {Ar{\'e}valo}, P., {Walton}, D.~J., {et~al.} 2015, \apj, 812,
  116

\bibitem[{{Brightman} {et~al.}(2017){Brightman}, {Balokovi{\'c}}, {Ballantyne},
  {Bauer}, {Boorman}, {Buchner}, {Brandt}, {Comastri}, {Del Moro}, {Farrah},
  {Gandhi}, {Harrison}, {Koss}, {Lanz}, {Masini}, {Ricci}, {Stern},
  {Vasudevan}, \& {Walton}}]{brig17ct}
{Brightman}, M., {Balokovi{\'c}}, M., {Ballantyne}, D.~R., {et~al.} 2017, \apj,
  844, 10

\bibitem[{{Cappelluti} {et~al.}(2017){Cappelluti}, {Li}, {Ricarte}, {Agarwal},
  {Allevato}, {Tasnim Ananna}, {Ajello}, {Civano}, {Comastri}, {Elvis},
  {Finoguenov}, {Gilli}, {Hasinger}, {Marchesi}, {Natarajan}, {Pacucci},
  {Treister}, \& {Urry}}]{capp17cosmos}
{Cappelluti}, N., {Li}, Y., {Ricarte}, A., {et~al.} 2017, \apj, 837, 19

\bibitem[{{Carroll} {et~al.}(2020){Carroll}, {Hickox}, {Masini}, {Lanz},
  {Assef}, {Stern}, {Chen}, \& {Ananna}}]{carr20}
{Carroll}, C.~M., {Hickox}, R.~C., {Masini}, A., {et~al.} 2020, arXiv e-prints,
  arXiv:2012.04668

\bibitem[{{Circosta} {et~al.}(2019){Circosta}, {Vignali}, {Gilli}, {Feltre},
  {Vito}, {Calura}, {Mainieri}, {Massardi}, \& {Norman}}]{circ19host}
{Circosta}, C., {Vignali}, C., {Gilli}, R., {et~al.} 2019, \aap, 623, A172

\bibitem[{{D'Amato} {et~al.}(2020){D'Amato}, {Gilli}, {Vignali}, {Massardi},
  {Pozzi}, {Zamorani}, {Circosta}, {Vito}, {Fritz}, {Cresci}, {Casasola},
  {Calura}, {Feltre}, {Manieri}, {Rigopoulou}, {Tozzi}, \&
  {Norman}}]{amat20alma}
{D'Amato}, Q., {Gilli}, R., {Vignali}, C., {et~al.} 2020, \aap, 636, A37

\bibitem[{{Diamond-Stanic} \& {Rieke}(2012)}]{dima12agnsf}
{Diamond-Stanic}, A.~M., \& {Rieke}, G.~H. 2012, \apj, 746, 168

\bibitem[{{Elbaz} {et~al.}(2011){Elbaz}, {Dickinson}, {Hwang},
  {D{\'\i}az-Santos}, {Magdis}, {Magnelli}, {Le Borgne}, {Galliano},
  {Pannella}, {Chanial}, {Armus}, {Charmandaris}, {Daddi}, {Aussel}, {Popesso},
  {Kartaltepe}, {Altieri}, {Valtchanov}, {Coia}, {Dannerbauer}, {Dasyra},
  {Leiton}, {Mazzarella}, {Alexander}, {Buat}, {Burgarella}, {Chary}, {Gilli},
  {Ivison}, {Juneau}, {Le Floc'h}, {Lutz}, {Morrison}, {Mullaney}, {Murphy},
  {Pope}, {Scott}, {Brodwin}, {Calzetti}, {Cesarsky}, {Charlot}, {Dole},
  {Eisenhardt}, {Ferguson}, {F{\"o}rster Schreiber}, {Frayer}, {Giavalisco},
  {Huynh}, {Koekemoer}, {Papovich}, {Reddy}, {Surace}, {Teplitz}, {Yun}, \&
  {Wilson}}]{elba11ms_aph}
{Elbaz}, D., {Dickinson}, M., {Hwang}, H.~S., {et~al.} 2011, \aap, 533, A119

\bibitem[{{Esquej} {et~al.}(2014){Esquej}, {Alonso-Herrero},
  {Gonz{\'a}lez-Mart{\'\i}n}, {H{\"o}nig}, {Hern{\'a}n-Caballero}, {Roche},
  {Ramos Almeida}, {Mason}, {D{\'\i}az-Santos}, {Levenson}, {Aretxaga},
  {Rodr{\'\i}guez Espinosa}, \& {Packham}}]{esqu14agnsf_aph}
{Esquej}, P., {Alonso-Herrero}, A., {Gonz{\'a}lez-Mart{\'\i}n}, O., {et~al.}
  2014, \apj, 780, 86

\bibitem[{{Fabbiano} {et~al.}(2017){Fabbiano}, {Elvis}, {Paggi}, {Karovska},
  {Maksym}, {Raymond}, {Risaliti}, \& {Wang}}]{fabb17fe}
{Fabbiano}, G., {Elvis}, M., {Paggi}, A., {et~al.} 2017, \apj, 842, L4

\bibitem[{{Freeman} {et~al.}(2001){Freeman}, {Doe}, \&
  {Siemiginowska}}]{free01sherpa}
{Freeman}, P., {Doe}, S., \& {Siemiginowska}, A. 2001, Society of Photo-Optical
  Instrumentation Engineers (SPIE) Conference Series, Vol. 4477, {Sherpa: a
  mission-independent data analysis application}, ed. J.-L. {Starck} \& F.~D.
  {Murtagh}, 76--87

\bibitem[{{Galametz} {et~al.}(2013){Galametz}, {Grazian}, {Fontana},
  {Ferguson}, {Ashby}, {Barro}, {Castellano}, {Dahlen}, {Donley}, {Faber},
  {Grogin}, {Guo}, {Huang}, {Kocevski}, {Koekemoer}, {Lee}, {McGrath}, {Peth},
  {Willner}, {Almaini}, {Cooper}, {Cooray}, {Conselice}, {Dickinson}, {Dunlop},
  {Fazio}, {Foucaud}, {Gardner}, {Giavalisco}, {Hathi}, {Hartley}, {Koo},
  {Lai}, {de Mello}, {McLure}, {Lucas}, {Paris}, {Pentericci}, {Santini},
  {Simpson}, {Sommariva}, {Targett}, {Weiner}, {Wuyts}, \& {CANDELS
  Team}}]{gala13candels}
{Galametz}, A., {Grazian}, A., {Fontana}, A., {et~al.} 2013, \apjs, 206, 10

\bibitem[{{Gallimore} {et~al.}(2016){Gallimore}, {Elitzur}, {Maiolino},
  {Marconi}, {O'Dea}, {Lutz}, {Baum}, {Nikutta}, {Impellizzeri}, {Davies},
  {Kimball}, \& {Sani}}]{gall16alma}
{Gallimore}, J.~F., {Elitzur}, M., {Maiolino}, R., {et~al.} 2016, \apjl, 829,
  L7

\bibitem[{{Gandhi} {et~al.}(2015){Gandhi}, {Yamada}, {Ricci}, {Asmus},
  {Mushotzky}, {Ueda}, {Terashima}, \& {La Parola}}]{gand15fe}
{Gandhi}, P., {Yamada}, S., {Ricci}, C., {et~al.} 2015, \mnras, 449, 1845

\bibitem[{{Georgantopoulos} \& {Akylas}(2019)}]{geor19ct}
{Georgantopoulos}, I., \& {Akylas}, A. 2019, \aap, 621, A28

\bibitem[{{Gilli} {et~al.}(2007){Gilli}, {Comastri}, \& {Hasinger}}]{gill07}
{Gilli}, R., {Comastri}, A., \& {Hasinger}, G. 2007, \aap, 463, 79

\bibitem[{{Gilli} {et~al.}(2014){Gilli}, {Norman}, {Vignali}, {Vanzella},
  {Calura}, {Pozzi}, {Massardi}, {Mignano}, {Casasola}, {Daddi}, {Elbaz},
  {Dickinson}, {Iwasawa}, {Maiolino}, {Brusa}, {Vito}, {Fritz}, {Feltre},
  {Cresci}, {Mignoli}, {Comastri}, \& {Zamorani}}]{gill14alma}
{Gilli}, R., {Norman}, C., {Vignali}, C., {et~al.} 2014, \aap, 562, A67

\bibitem[{{Guo} {et~al.}(2013){Guo}, {Ferguson}, {Giavalisco}, {Barro},
  {Willner}, {Ashby}, {Dahlen}, {Donley}, {Faber}, {Fontana}, {Galametz},
  {Grazian}, {Huang}, {Kocevski}, {Koekemoer}, {Koo}, {McGrath}, {Peth},
  {Salvato}, {Wuyts}, {Castellano}, {Cooray}, {Dickinson}, {Dunlop}, {Fazio},
  {Gardner}, {Gawiser}, {Grogin}, {Hathi}, {Hsu}, {Lee}, {Lucas}, {Mobasher},
  {Nand ra}, {Newman}, \& {van der Wel}}]{guo13candels}
{Guo}, Y., {Ferguson}, H.~C., {Giavalisco}, M., {et~al.} 2013, \apjs, 207, 24

\bibitem[{{Hickox} \& {Alexander}(2018)}]{hick18}
{Hickox}, R.~C., \& {Alexander}, D.~M. 2018, Annual Review of Astronomy and
  Astrophysics, 56, 625

\bibitem[{{Hickox} {et~al.}(2007){Hickox}, {Jones}, {Forman}, {Murray},
  {Brodwin}, {Brown}, {Eisenhardt}, {Stern}, {Kochanek}, {Eisenstein}, {Cool},
  {Jannuzi}, {Dey}, {Brand}, {Gorjian}, \& {Caldwell}}]{hick07abs}
{Hickox}, R.~C., {Jones}, C., {Forman}, W.~R., {et~al.} 2007, \apj, 671, 1365

\bibitem[{{Iwasawa} {et~al.}(2012){Iwasawa}, {Mainieri}, {Brusa}, {Comastri},
  {Gilli}, {Vignali}, {Hasinger}, {Sanders}, {Cappelluti}, {Impey},
  {Koekemoer}, {Lanzuisi}, {Lusso}, {Merloni}, {Salvato}, {Taniguchi}, \&
  {Trump}}]{iwas12fe}
{Iwasawa}, K., {Mainieri}, V., {Brusa}, M., {et~al.} 2012, \aap, 537, A86

\bibitem[{{Iwasawa} {et~al.}(2020){Iwasawa}, {Comastri}, {Vignali}, {Gilli},
  {Lanzuisi}, {Brandt}, {Tozzi}, {Brusa}, {Carrera}, {Ranalli}, {Mainieri},
  {Georgantopoulos}, {Puccetti}, \& {Paolillo}}]{iwas20cdfs}
{Iwasawa}, K., {Comastri}, A., {Vignali}, C., {et~al.} 2020, \aap, 639, A51

\bibitem[{{Jones} {et~al.}(2020){Jones}, {Fabbiano}, {Elvis}, {Paggi},
  {Karovska}, {Maksym}, {Siemiginowska}, \& {Raymond}}]{jone20fe}
{Jones}, M.~L., {Fabbiano}, G., {Elvis}, M., {et~al.} 2020, \apj, 891, 133

\bibitem[{{Kennicutt}(1998)}]{kenn98araa}
{Kennicutt}, Jr., R.~C. 1998, \araa, 36, 189

\bibitem[{{LaMassa} {et~al.}(2011){LaMassa}, {Heckman}, {Ptak}, {Martins},
  {Wild}, {Sonnentrucker}, \& {Hornschemeier}}]{lama11fe}
{LaMassa}, S.~M., {Heckman}, T.~M., {Ptak}, A., {et~al.} 2011, \apj, 729, 52

\bibitem[{{Lambrides} {et~al.}(2020){Lambrides}, {Chiaberge}, {Heckman},
  {Gilli}, {Vito}, \& {Norman}}]{lamb20cdfs}
{Lambrides}, E.~L., {Chiaberge}, M., {Heckman}, T., {et~al.} 2020, \apj, 897,
  160

\bibitem[{{Lansbury} {et~al.}(2015){Lansbury}, {Gandhi}, {Alexander}, {Assef},
  {Aird}, {Annuar}, {Ballantyne}, {Balokovi{\'c}}, {Bauer}, {Boggs}, {Brandt},
  {Brightman}, {Christensen}, {Civano}, {Comastri}, {Craig}, {Del Moro},
  {Grefenstette}, {Hailey}, {Harrison}, {Hickox}, {Koss}, {LaMassa}, {Luo},
  {Puccetti}, {Stern}, {Treister}, {Vignali}, {Zappacosta}, \&
  {Zhang}}]{lans15nustarqso}
{Lansbury}, G.~B., {Gandhi}, P., {Alexander}, D.~M., {et~al.} 2015, \apj, 809,
  115

\bibitem[{{Lanzuisi} {et~al.}(2018){Lanzuisi}, {Civano}, {Marchesi},
  {Comastri}, {Brusa}, {Gilli}, {Vignali}, {Zamorani}, {Brightman},
  {Griffiths}, \& {Koekemoer}}]{lanz18ct}
{Lanzuisi}, G., {Civano}, F., {Marchesi}, S., {et~al.} 2018, \mnras, 480, 2578

\bibitem[{{Levenson} {et~al.}(2002){Levenson}, {Krolik}, {{\.Z}ycki},
  {Heckman}, {Weaver}, {Awaki}, \& {Terashima}}]{leve02fe}
{Levenson}, N.~A., {Krolik}, J.~H., {{\.Z}ycki}, P.~T., {et~al.} 2002, \apjl,
  573, L81

\bibitem[{{Luo} {et~al.}(2017){Luo}, {Brandt}, {Xue}, {Lehmer}, {Alexander},
  {Bauer}, {Vito}, {Yang}, {Basu-Zych}, {Comastri}, {Gilli}, {Gu},
  {Hornschemeier}, {Koekemoer}, {Liu}, {Mainieri}, {Paolillo}, {Ranalli},
  {Rosati}, {Schneider}, {Shemmer}, {Smail}, {Sun}, {Tozzi}, {Vignali}, \&
  {Wang}}]{luo17chandra}
{Luo}, B., {Brandt}, W.~N., {Xue}, Y.~Q., {et~al.} 2017, \apjs, 228, 2

\bibitem[{{Marchesi} {et~al.}(2018){Marchesi}, {Ajello}, {Marcotulli},
  {Comastri}, {Lanzuisi}, \& {Vignali}}]{marc18ct}
{Marchesi}, S., {Ajello}, M., {Marcotulli}, L., {et~al.} 2018, \apj, 854, 49

\bibitem[{{Marinucci} {et~al.}(2017){Marinucci}, {Bianchi}, {Fabbiano}, {Matt},
  {Risaliti}, {Nardini}, \& {Wang}}]{mari17fe}
{Marinucci}, A., {Bianchi}, S., {Fabbiano}, G., {et~al.} 2017, \mnras, 470,
  4039

\bibitem[{{Mateos} {et~al.}(2017){Mateos}, {Carrera}, {Barcons},
  {Alonso-Herrero}, {Hern{\'a}n-Caballero}, {Page}, {Ramos Almeida},
  {Caccianiga}, {Miyaji}, \& {Blain}}]{mate17obsc}
{Mateos}, S., {Carrera}, F.~J., {Barcons}, X., {et~al.} 2017, \apjl, 841, L18

\bibitem[{{Matt} {et~al.}(1991){Matt}, {Perola}, \& {Piro}}]{matt91}
{Matt}, G., {Perola}, G.~C., \& {Piro}, L. 1991, \aap, 247, 25

\bibitem[{{Mullaney} {et~al.}(2011){Mullaney}, {Alexander}, {Goulding}, \&
  {Hickox}}]{mull11agnsed}
{Mullaney}, J.~R., {Alexander}, D.~M., {Goulding}, A.~D., \& {Hickox}, R.~C.
  2011, \mnras, 414, 1082

\bibitem[{{Mullaney} {et~al.}(2012){Mullaney}, {Pannella}, {Daddi}, {Alexand
  er}, {Elbaz}, {Hickox}, {Bournaud}, {Altieri}, {Aussel}, {Coia},
  {Dannerbauer}, {Dasyra}, {Dickinson}, {Hwang}, {Kartaltepe}, {Leiton},
  {Magdis}, {Magnelli}, {Popesso}, {Valtchanov}, {Bauer}, {Brand t}, {Del
  Moro}, {Hanish}, {Ivison}, {Juneau}, {Luo}, {Lutz}, {Sargent}, {Scott}, \&
  {Xue}}]{mull12fir}
{Mullaney}, J.~R., {Pannella}, M., {Daddi}, E., {et~al.} 2012, \mnras, 419, 95

\bibitem[{{Netzer}(2015)}]{netz15unified}
{Netzer}, H. 2015, \araa, 53, 365

\bibitem[{{Panagiotou} \& {Walter}(2019)}]{pana19torus}
{Panagiotou}, C., \& {Walter}, R. 2019, \aap, 626, A40

\bibitem[{{Park} {et~al.}(2006){Park}, {Kashyap}, {Siemiginowska}, {van Dyk},
  {Zezas}, {Heinke}, \& {Wargelin}}]{park06behr}
{Park}, T., {Kashyap}, V.~L., {Siemiginowska}, A., {et~al.} 2006, \apj, 652,
  610

\bibitem[{{P{\'e}rez-Torres} {et~al.}(2021){P{\'e}rez-Torres}, {Mattila},
  {Alonso-Herrero}, {Aalto}, \& {Efstathiou}}]{pere20lirg}
{P{\'e}rez-Torres}, M., {Mattila}, S., {Alonso-Herrero}, A., {Aalto}, S., \&
  {Efstathiou}, A. 2021, \aapr, 29, 2

\bibitem[{{Ricci} {et~al.}(2015){Ricci}, {Ueda}, {Koss}, {Trakhtenbrot},
  {Bauer}, \& {Gandhi}}]{ricc15ct}
{Ricci}, C., {Ueda}, Y., {Koss}, M.~J., {et~al.} 2015, \apjl, 815, L13

\bibitem[{{Ricci} {et~al.}(2014){Ricci}, {Ueda}, {Paltani}, {Ichikawa},
  {Gandhi}, \& {Awaki}}]{ricc14fe}
{Ricci}, C., {Ueda}, Y., {Paltani}, S., {et~al.} 2014, \mnras, 441, 3622

\bibitem[{{Salpeter}(1955)}]{salp55lf}
{Salpeter}, E.~E. 1955, \apj, 121, 161

\bibitem[{{Santini} {et~al.}(2015){Santini}, {Ferguson}, {Fontana}, {Mobasher},
  {Barro}, {Castellano}, {Finkelstein}, {Grazian}, {Hsu}, {Lee}, {Lee},
  {Pforr}, {Salvato}, {Wiklind}, {Wuyts}, {Almaini}, {Cooper}, {Galametz},
  {Weiner}, {Amorin}, {Boutsia}, {Conselice}, {Dahlen}, {Dickinson},
  {Giavalisco}, {Grogin}, {Guo}, {Hathi}, {Kocevski}, {Koekemoer},
  {Kurczynski}, {Merlin}, {Mortlock}, {Newman}, {Paris}, {Pentericci},
  {Simons}, \& {Willner}}]{sant15sed}
{Santini}, P., {Ferguson}, H.~C., {Fontana}, A., {et~al.} 2015, \apj, 801, 97

\bibitem[{{Shu} {et~al.}(2010){Shu}, {Yaqoob}, \& {Wang}}]{shu10fe}
{Shu}, X.~W., {Yaqoob}, T., \& {Wang}, J.~X. 2010, \apjs, 187, 581

\bibitem[{{Speagle} {et~al.}(2014){Speagle}, {Steinhardt}, {Capak}, \&
  {Silverman}}]{spea14ms}
{Speagle}, J.~S., {Steinhardt}, C.~L., {Capak}, P.~L., \& {Silverman}, J.~D.
  2014, \apjs, 214, 15

\bibitem[{{Stark} {et~al.}(1992){Stark}, {Gammie}, {Wilson}, {Bally}, {Linke},
  {Heiles}, \& {Hurwitz}}]{star92}
{Stark}, A.~A., {Gammie}, C.~F., {Wilson}, R.~W., {et~al.} 1992, \apjs, 79, 77

\bibitem[{{Trebitsch} {et~al.}(2019){Trebitsch}, {Volonteri}, \&
  {Dubois}}]{treb19}
{Trebitsch}, M., {Volonteri}, M., \& {Dubois}, Y. 2019, \mnras, 487, 819

\bibitem[{{Treister} {et~al.}(2009){Treister}, {Urry}, \& {Virani}}]{trei09}
{Treister}, E., {Urry}, C.~M., \& {Virani}, S. 2009, \apj, 696, 110

\bibitem[{{Ueda} {et~al.}(2014){Ueda}, {Akiyama}, {Hasinger}, {Miyaji}, \&
  {Watson}}]{ueda14cxb}
{Ueda}, Y., {Akiyama}, M., {Hasinger}, G., {Miyaji}, T., \& {Watson}, M.~G.
  2014, \apj, 786, 104

\bibitem[{{Wright}(2006)}]{wrig06coscal}
{Wright}, E.~L. 2006, \pasp, 118, 1711

\bibitem[{{Yan} {et~al.}(2019){Yan}, {Hickox}, {Hainline}, {Stern}, {Lansbury},
  {Alexander}, {Hviding}, {Assef}, {Ballantyne}, {Dipompeo}, {Lanz}, {Carroll},
  {Koss}, {Lamperti}, {Civano}, {Del Moro}, {Gandhi}, \& {Myers}}]{yan19nustar}
{Yan}, W., {Hickox}, R.~C., {Hainline}, K.~N., {et~al.} 2019, \apj, 870, 33

\bibitem[{{Yang} {et~al.}(2017){Yang}, {Chen}, {Vito}, {Brandt}, {Alexander},
  {Luo}, {Sun}, {Xue}, {Bauer}, {Koekemoer}, {Lehmer}, {Liu}, {Schneider},
  {Shemmer}, {Trump}, {Vignali}, \& {Wang}}]{yang17}
{Yang}, G., {Chen}, C. T.~J., {Vito}, F., {et~al.} 2017, \apj, 842, 72

\end{thebibliography}

 \newcommand{\noop}[1]{}

\end{document}